\documentclass[prl, twocolumn, superscriptaddress, longbibliography]{revtex4-1}

\usepackage{amssymb}
\usepackage{amsmath}
\usepackage{graphicx}
\usepackage{amsfonts}
\usepackage[usenames, dvipsnames]{xcolor}
\usepackage{bm}
\usepackage{bbm}
\usepackage{color}
\usepackage{hyperref}
\hypersetup{pdfnewwindow=true,
            colorlinks=true,
            linkcolor=blue,
            citecolor=blue,
            urlcolor=blue}
\usepackage{soul}
\usepackage[caption=false, justification=centerlast]{subfig}
\usepackage{psfrag}
\usepackage{textcomp}
\usepackage{braket}

\DeclareMathOperator{\Tr}{Tr}
\DeclareMathOperator{\Var}{Var}

\newcommand{\F}{\mathcal{F}}

\begin{document}

\title{Thermometric machine for ultraprecise thermometry of low temperatures}

\author{Ivan Henao}
\email{ivan.henao@mail.huji.ac.il}
\affiliation{Fritz Haber Research Center for Molecular Dynamics, Institute of Chemistry, The Hebrew University of Jerusalem, Jerusalem 9190401, Israel}

\author{Karen V. Hovhannisyan}
\email{karen.hovhannisyan@uni-potsdam.de}
\affiliation{The Abdus Salam International Centre for Theoretical Physics (ICTP), Strada Costiera 11, 34151 Trieste, Italy}
\affiliation{Institute of Physics and Astronomy, University of Potsdam, 14476 Potsdam, Germany}

\author{Raam Uzdin}
\email{raam@mail.huji.ac.il}
\affiliation{Fritz Haber Research Center for Molecular Dynamics, Institute of Chemistry, The Hebrew University of Jerusalem, Jerusalem 9190401, Israel}

\begin{abstract}
Thermal equilibrium states are exponentially hard to distinguish at very low temperatures, making equilibrium quantum thermometry in this regime a formidable task. We present a thermometric scheme that circumvents this limitation, by using a two-level probe that does not thermalize with the sample whose temperature is measured. This is made possible thanks to a suitable interaction that couples the probe to the sample and to an auxiliary thermal bath known to be at a higher temperature. Provided a reasonable upper bound on the temperature of the sample, the resulting “thermometric machine” drives the probe towards a steady state whose signal-to-noise ratio can achieve values as high as $\mathcal{O}(1/T)$. We also characterize the transient state of the probe and numerically illustrate an extreme reduction in the number of measurements to attain a given precision, as compared to optimal measurements on a thermalized probe. \end{abstract}

\maketitle

Estimating the temperature of a quantum system is a task of great importance in fundamental and applied research. In the study of fundamental quantum phenomena and the development of related applications, it is crucial to precisely prepare and characterize systems at temperatures near the absolute zero. For example, quantum computing based on superconducting qubits operates with temperatures of the order of mK \cite{krinner2019engineering}. In addition, a leading platform for the study of condensed matter physics and with great potential for quantum simulation are ultracold atoms, where temperatures as low as picokelvins have been achieved \cite{chen2020emergence}. In the last years, these and other applications have spurred an increasing theoretical \cite{stace2010quantum, jahnke2011operational, correa2015individual, paris2015achieving, mehboudi2015thermometry, pasquale2016local, hofer2017quantum, correa2017enhancement, hovhannisyan2018measuring, plodzien2018few, potts2019fundamental, mehboudi2019using, mukherjee2019enhanced, mehboudi2019thermometry, jorgensen2020tight, mitchison2020situ} and experimental \cite{giazotto2006opportunities, feshchenko2015tunnel, halbertal2016nanoscale, iftikhar2016primary, karimi2018noninvasive, mehboudi2019thermometry, scigliuzzo2020primary} effort to understand and improve the performance of low-temperature thermometry.

Unfortunately, the temperature encoded in a system at thermal equilibrium becomes harder to measure the lower it is. For a quantum system characterized by a Hamiltonian $H=\sum_{i}\varepsilon_{i}|i\rangle\langle i|$, with eigenenergies $\{\varepsilon_{i}\}$, the temperature regime where this difficulty arises is determined by the condition $T\ll\varepsilon_{1}-\varepsilon_{0}$ \cite{paris2015achieving, potts2019fundamental, mehboudi2019thermometry}. 
Accordingly, temperatures much lower than the energy gap above the ground state present a significant challenge for quantum thermometry \cite{paris2015achieving, correa2017enhancement, hovhannisyan2018measuring, potts2019fundamental, mehboudi2019thermometry}. As discussed later, this is a consequence of the Cram\'{e}r-Rao bound 
\begin{equation} \label{eq:1 CRB for SNR}
\frac{T}{\Delta T} \leq T \sqrt{M \F(T)},
\end{equation}
which poses a fundamental limit to the thermometric precision \cite{ mehboudi2019thermometry}. Here, the quantity $T/\Delta T$ is known as the signal-to-noise ratio (SNR), and $\Delta T$ is the absolute error in the estimation of $T$. While the bound (\ref{eq:1 CRB for SNR}) is typically expressed in the equivalent form $\Delta T\geq1/\sqrt{M\mathcal{F}(T)}$, the SNR is in our opininion a better quantifier of precision. Specifically, a large SNR implies that the error $\Delta T$ is small as compared to the 'signal' $T$. 

The key element at the r.h.s. of Eq. (\ref{eq:1 CRB for SNR}) is the Fisher information $\mathcal{F}(T)$ \cite{paris2009quantum}. Given a temperature-dependent quantum state $\rho(T)$, the most general strategy to estimate $T$ consists of performing a POVM (positive operator valued measurement) on $\rho(T)$, followed by a classical postprocessing that asigns temperature values to the different POVM outcomes. If the $j$th outcome occurs with probability $p_{j}(T)$ the Fisher information reads \cite{cramer1946mathematical, braunstein1994statistical}
\begin{equation}\label{eq:1.1 Fisher information}
	\mathcal{F}(T)=\sum_{j}p_{j}(T)\left(\frac{\partial\textrm{ln}(p_{j})}{\partial T}\right)^{2},
\end{equation}
where the sum runs over all the POVM outcomes. In this way, Eq. (\ref{eq:1 CRB for SNR}) limits the SNR for a temperature estimate drawn from $M$ independent realisations of the POVM. For a thermal state $\rho(T)=\rho_{\textrm{th}}(T)=\frac{e^{-\beta H}}{\textrm{Tr}\left(e^{-\beta H}\right)}$ (where $\beta=1/T$ is the inverse temperature and we set the Boltzmann constant equal to one), the POVM that maximizes $\mathcal{F}(T)$ corresponds to projective measurements of the Hamiltonian $H$. In such a case the bound (\ref{eq:1 CRB for SNR}) is also tight, for any value of $M$ \cite{ mehboudi2019thermometry}. 

In many-body systems, temperature measurements via small probes \cite{hovhannisyan2018measuring,hovhannisyan2021optimal,correa2015individual,mitchison2020situ,jevtic2015single,gebbia2020two,brunelli2011qubit,brunelli2012qubit,razavian2019quantum,sekatski2021optimal,correa2017enhancement} constitute a more practical or convenient approach for thermometry, due to the destructive character or practical difficulty of direct measurements \cite{mehboudi2015thermometry,mitchison2020situ,mehboudi2019using,hangleiter2015nondestructive,hovhannisyan2021optimal}. However, probes that thermalize with the system of interest (the 'sample') are rather limited in their probing capabilities  \cite{hovhannisyan2018measuring,paris2015achieving,correa2015individual,potts2019fundamental}, in line with our initial observations. To understand this better consider the Fisher information $\mathcal{F}_{P}^{th}(T)$, for energy measurements on a thermalized (hence the superscript '$th$') probe. In what follows, the indices $P$ and $S$ will refer to the probe and the sample, respectively. Given an infinitesimal temperature variation $dT$, $\mathcal{F}_{P}^{th}(T)$ quantifies the degree of distinguishability between two states $\rho_{P}^{th}(T)$ and $\rho_{P}^{th}(T+dT)$ \cite{correa2015individual,braunstein1994statistical}. In the limit $T\rightarrow0$, this distinguishability is suppressed by an exponential scaling $\mathcal{F}_{P}^{th}(T)\sim\mathcal{O}\left(e^{-\beta(\varepsilon_{1}^{P}-\varepsilon_{0}^{P})}\right)$
\cite{paris2015achieving,pasquale2016local,brandao2015entanglement,correa2017enhancement,hovhannisyan2018measuring,potts2019fundamental} that also affects the SNR (cf. Eq. \eqref{eq:1 CRB for SNR}). Therefore, for $T\ll\varepsilon_{1}^{P}-\varepsilon_{0}^{P}$ thermometry using a (finite) thermal probe is exponentially inefficient.

If the sample itself is gapped (i.e. $\varepsilon_{1}^{S}-\varepsilon_{0}^{S}$ is finite), the exponential decay occurs even for probes that do not couple weakly to the sample and thus attain a non-Gibbsian state when equilibrated with the sample \cite{hovhannisyan2018measuring}. A gapless sample is characterized by a continuous spectrum above the ground state, which implies particularly that $\varepsilon_{1}^{S}\rightarrow\varepsilon_{0}^{S}$. The precision of a probe that strongly couples to such a sample is not exponentially suppressed anymore; at equilibrium, a generic probe will deliver only a polynomially decaying (with respect to $\beta$) SNR \cite{hovhannisyan2018measuring,potts2019fundamental}. In gapless harmonic systems, a free Brownian particle, when used as a probe, can yield an SNR that is as high as $\mathcal{O}(1)$ \cite{hovhannisyan2018measuring}. The latter is the highest temperature scaling reported for equilibrium probe-based thermometry \footnote{We emphasize that equilibrium is understood globally. While a strongly interacting probe is described by a non-Gibbsian state, the global state (including the sample) is a thermal equilibrium state.}.

Motivated by the observation that non-equilibrium probes can provide better temperature sensing \cite{mukherjee2019enhanced,hovhannisyan2021optimal,henao2020catalytic,seah2019collisional}, in this letter we design a non-equilibrium thermometric scheme that allows us to achieve a \textit{divergent} SNR in the limit $T\rightarrow0$. The key element behind this improvement is the use of an ancillary system \cite{kiilerich2018dynamical} that interacts with the probe and the sample, and builds upon the recent finding that probe-based thermometry can be catalytically enhanced \cite{henao2020catalytic}. These interactions are sequentially repeated with different sample constituents (see Fig. 1(a)), in the spirit of a collisional model \cite{campbell2021collision,seah2019collisional,ciccarello2021quantum,shu2020surpassing}. Crucially, the resulting evolution drives the probe to a non-thermal steady state, with a dependence on ancillary parameters that can be tuned to suppress the exponential scaling characteristic of low temperatures. Moreover, this mitigation can be enforced without any restriction on the energy gaps of the sample. 

\begin{figure}
\includegraphics[scale=0.6]{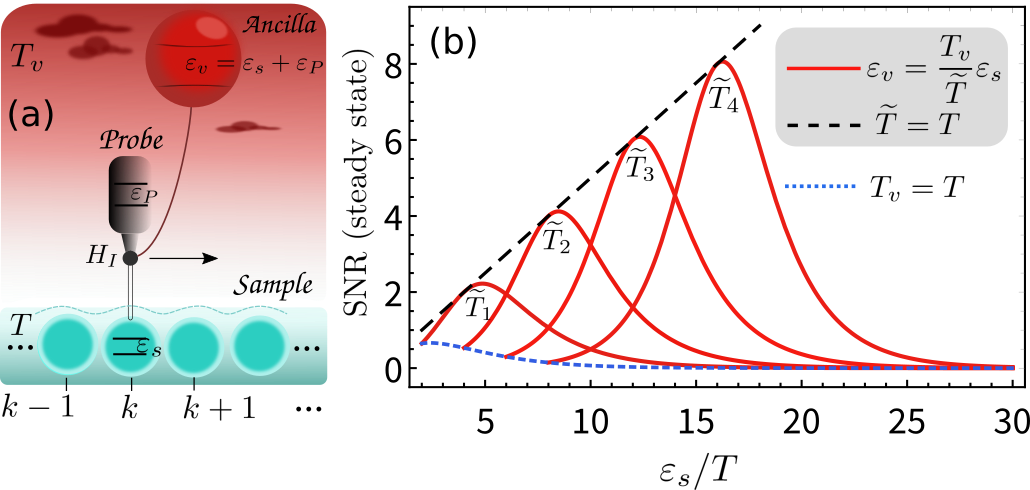}
\caption{(a) Thermometric machine. A two-level system with energy gap $\varepsilon_{P}$ probes thermal qubits at temperature $T$ (with energy gap $\varepsilon_{s}$), through sequential interactions that include also a thermal ancilla at temperature $T_{v}>T$ (with energy gap $\varepsilon_{v}=\varepsilon_{s}+\varepsilon_{P}$).
(b) Steady-state signal-to-noise ratio (SNR) of the probe. From left to right, the red solid curves give the SNRs obtained for temperatures \textit{known to belong to intervals} $(0,\varepsilon_{s}/2)$, $(0,\varepsilon_{s}/4)$, $(0,\varepsilon_{s}/6)$, and $(0,\varepsilon_{s}/8)$. The labels $\widetilde{T}_{i}$ correspond to \textit{prior} \textit{temperatures} that characterize this knowledge (cf. (\ref{eq:4 Tprior})). The black dashed line is the SNR corresponding to $T=\widetilde{T}$, and the blue (dotted) curve depicts the thermal SNR of the probe [cf. Eq.~\eqref{eq:8 Thermal SNR}], assuming $\varepsilon_{P}=\varepsilon_{s}$. In all the plots $M=1$ is assumed.} 
\end{figure}

\textbf{\textit{Thermometric machine}}. For the sake of simplicity, we will consider a sample composed of $N$ qubits. However, the results presented later for the steady state of the probe are generalizable to samples formed by $N$ identical and non-interacting systems, of arbitrary dimension and energy spectrum (see Sec. I of Supplemental Material (SM)
\footnote{See Supplemental Material at URL, which includes Refs.}).
The probe and the ancilla are two-level systems with Hamiltonians $H_{P}=\varepsilon_{P}|1_{P}\rangle\langle1_{P}|$
and $H_{v}=\varepsilon_{v}|1_{v}\rangle\langle1_{v}|$, respectively.
Moreover, each sample qubit has Hamiltonian $H_{s}=\varepsilon_{s}|1_{s}\rangle\langle1_{s}|$, and $H_{S}=\sum_{i=1}^{N}H_{s}^{(i)}$ is the total sample Hamiltonian. For a sample prepared in the thermal state $\rho_{S}=\rho_{s}^{\otimes N}$, where $\rho_{s}=\sum_{i=0}^{1}p_{i}^{s}|i_{s}\rangle\langle i_{s}|=\frac{e^{-\beta H_{s}}}{\Tr\left(e^{-\beta H_{s}}\right)}$, temperature information is transferred to the probe in the following way:

\textbf{(1)} After preparing the ancilla in a thermal state $\rho_{v}=\sum_{i=0}^{1}p_{i}^{v}|i_{v}\rangle\langle i_{v}|=\frac{e^{-\beta_{v}H_{v}}}{\Tr\left(e^{-\beta_{v} H_{v}}\right)}$, where $\beta_{v}<\beta$, it is coupled to the probe and to a single sample qubit via the three-body interaction $H_{I}=\varepsilon_{I}(|0_{P}0_{s}1_{v}\rangle\langle1_{P}1_{s}0_{v}|+\textrm{h.c.})$ (where h.c. stands for Hermitian conjugate and $|i_{P}j_{s}k_{v}\rangle=|i_{P}\rangle\otimes|j_{s}\rangle\otimes|k_{v}\rangle$). In addition, we assume that $\varepsilon_{v}=\varepsilon_{P}+\varepsilon_{s}$, which implies 
\begin{equation}
[H_{I},H_{P}+H_{s}+H_{v}]=0.\label{eq:2 commutation relation}
\end{equation}

\textbf{(2)} The interaction $H_{I}$ is switched off and the ancilla is rethermalized.

Each repetion of Step \textbf{(1)} gives rise to a “collision” between the probe, the ancilla, and the sample \cite{campbell2021collision}. If $H_{I}$ is switched on during a time $\pi/2\varepsilon_{I}$, the total Hamiltonian $H_{Psv}=H_{P}+H_{s}+H_{v}+H_{I}$ yields the unitary evolution (apart from unimportant global phases, see Sec. II of SM \cite{Note2} for further details) 
\begin{align}
U_{Psv}|0_{P}0_{s}1_{v}\rangle & =|1_{P}1_{s}0_{v}\rangle,\label{eq:3.1 U_Psv}\\
U_{Psv}|1_{P}1_{s}0_{v}\rangle & =|0_{P}0_{s}1_{v}\rangle.\label{eq:3.2 U_Psv}
\end{align}
Furthermore, $U_{Psv}$ acts as the identity on any other eigenstate
of $H_{P}+H_{s}+H_{v}$. 

\textbf{\textit{Tuning of}} $\varepsilon_{v}$. Since $\varepsilon_{s}$ is the energy gap above the ground state of the sample, the corresponding low-temperature regime can be characterized by temperatures $T$ in an interval $(0,T_{\textrm{max}})$ such that $T_{\textrm{max}}\ll\varepsilon_{s}$. Given a fixed temperature $T_{v}$, this information can be harnessed to properly choose the energy gap $\varepsilon_{v}$. To that end, we consider the \textit{prior} temperature 
\begin{equation} \label{eq:4 Tprior}
\widetilde{T}\equiv\frac{T_{\textrm{max}}}{2},
\end{equation}
which encapsulates the knowledge that $T\in(0,T_{\textrm{max}})$. In this way, the goal of the thermometric machine is to estimate with as much precision as possible the specific value of $T$. 

Let $\rho_{P}^{(k)}(T)$ denote the state of the probe after it has interacted with $k$ qubits, and let $T/\Delta_{k}T$ be the SNR that results from $M$ (independent) energy measurements performed on $\rho_{P}^{(k)}(T)$. In Sec. II of SM \cite{Note2} we show that this state is diagonal in the eigenbasis of the probe Hamiltonian, i.e. $\rho_{P}^{(k)}(T)=\sum_{i=0}^{1}p_{i,k}^{P}(T)|i_{P}\rangle\langle i_{P}|$. Accordingly, the Fisher information and the corresponding Cramer-Rao bound for the SNR are maximized by performing energy measurements. 

The statistics obtained from a two-level system belong to the so-called exponential family \cite{vanderVaart} and therefore the associated Cram\'{e}r-Rao bound is tight. For the sake of completeness, we also present a detailed proof based on the error-propagation formula in Sec. III of SM \cite{Note2}.
Hence, the \textit{exact} steady-state SNR of the probe, defined by the limits $N\rightarrow\infty$ and $k\rightarrow\infty$, is given by 

\begin{align} \label{eq:5 steady SNR}
\frac{T}{\Delta_{\infty}T} & =\sqrt{M\left(p_{0,\infty}^{P}(T)p_{1,\infty}^{P}(T)\right)}\frac{\varepsilon_{s}}{T}\nonumber \\
 & =\frac{\sqrt{M}e^{-\frac{1}{2}(\beta\varepsilon_{s}-\beta_{v}\varepsilon_{v})}}{1+e^{-(\beta\varepsilon_{s}-\beta_{v}\varepsilon_{v})}}\frac{\varepsilon_{s}}{T}=\frac{\sqrt{M}e^{-\frac{1}{2}\beta\varepsilon_{s}\left(1-\frac{2T}{T_{\textrm{max}}}\right)}}{1+e^{-\beta\varepsilon_{s}\left(1-\frac{2T}{T_{\textrm{max}}}\right)}}\frac{\varepsilon_{s}}{T},
\end{align}
where the prior temperature $\widetilde{T}$ is used to tune the ancillary energy gap as
\begin{equation} \label{eq:6 ancillary tuning}
\varepsilon_{v}=\frac{T_{v}}{\widetilde{T}}\varepsilon_{s}=2\frac{T_{v}}{T_{\textrm{max}}}\varepsilon_{s}.
\end{equation}

Assuming that $T_{v}\geq T_{\textrm{max}}$, Eq. (\ref{eq:6 ancillary tuning})
implies that $\varepsilon_{v}\geq2\varepsilon_{s}$, and from the
condition $\varepsilon_{P}=\varepsilon_{v}-\varepsilon_{s}$ it also
follows that $\varepsilon_{P}\geq\varepsilon_{s}$. Since this implies
that $\varepsilon_{s}=\textrm{min}\{\varepsilon_{s},\varepsilon_{v},\varepsilon_{P}\}$,
for the tuning (\ref{eq:6 ancillary tuning}) and $T_{v}\geq T_{\textrm{max}}$
precise estimation of temperatures in the ultracold regime $T\ll\varepsilon_{s}$
is possible even if the energy gaps of the ancilla and the probe are
arbitrarily larger than $T$. Hence, \textit{the advantage provided
by the machine does not involve any gapless system}. 

\textbf{\textit{Steady-state regime}}. The red curves in Fig. 1(b)
depict steady-state SNRs for $M=1$ and energy gaps $\varepsilon_{v}$ that
satisfy Eq. (\ref{eq:6 ancillary tuning}). The prior temperatures
$\widetilde{T}$ labeling each curve are given by $\widetilde{T}_{1}=\varepsilon_{s}/4$,
$\widetilde{T}_{2}=\varepsilon_{s}/8$, $\widetilde{T}_{3}=\varepsilon_{s}/12$,
and $\widetilde{T}_{4}=\varepsilon_{s}/16$. Hence, the respective
tasks consist of estimating temperatures that belong to the intervals
$(0,\varepsilon_{s}/2)$, $(0,\varepsilon_{s}/4)$, $(0,\varepsilon_{s}/6)$,
and $(0,\varepsilon_{s}/8)$. The red curves intersect the black dashed
line at $T=\widetilde{T}$, where $T=T_{\textrm{max}}/2$. According
to Eqs. (\ref{eq:4 Tprior}) and (\ref{eq:5 steady SNR}), in this
case 
\begin{equation}
\frac{T}{\Delta_{\infty}T}=\frac{1}{2}\frac{\varepsilon_{s}}{T}=\frac{1}{2}\frac{\varepsilon_{s}}{\widetilde{T}}.\label{eq:7 Maximum steady SNR}
\end{equation}
Therefore, for $T=\widetilde{T}$ the SNR \textit{diverges} as $T$ tends to zero. 

As we show in Sec. VI of SM \cite{Note2}, the scaling $\mathcal{O}(\varepsilon_{s}/\widetilde{T})$ in Eq.~\eqref{eq:7 Maximum steady SNR} is robust under the practical constraint of a not perfectly known temperature $T_{v}$. Specifically, in the interval $T\in(0,T_{\textrm{max}})$ the steady-state SNR can
attain a value $\frac{1}{2}\left(1\pm\frac{\Delta T_{v}}{T_{v}}\right)\frac{\varepsilon_{s}}{\tilde{T}}$
if the relative error in the estimation of $T_{v}$ satisfies $\Delta T_{v}/T_{v}\leq1/2$.
Since $T_{v}$ can be much larger than $T$ the temperature $T_{v}$ can be efficiently estimated, e.g., by performing energy measurements on a thermalized two-level system \footnote{In particular, energy measurements on a thermal two-level system with energy gap $\approx 2.5 T_v$ yield $\Delta T_v/T_v \approx 0.4 < 1/2$ for $M = 16$ measurements; see Sec. VI of SM \cite{Note1}.}.

The blue dotted curve in Fig. 1(b) shows the thermal SNR, obtained
for $T_{v}=T$. In this case, the probe equilibrates to the thermal
state associated with $T$. This is not surprising, since $H_{I}$
commutes with the total free Hamiltonian $H_{P}+H_{s}+H_{v}$ (cf.
Eq. (\ref{eq:2 commutation relation})) and therefore the thermal
state $\rho_{Psv}(T)=\frac{e^{-\beta(H_{s}+H_{P}+H_{v})}}{\Tr\left[e^{-\beta(H_{s}+H_{P}+H_{v})}\right]}$
is invariant under $U_{Psv}$. By applying Eq. (\ref{eq:1 CRB for SNR})
to $\rho_{P}(T)=\Tr_{sv}\left[\rho_{Psv}(T)\right]$ and assuming
$\varepsilon_{P}=\varepsilon_{s}$, it follows that 
\begin{equation}
\frac{T}{\Delta T}=\sqrt{M}\frac{e^{-\frac{1}{2}\beta\varepsilon_{s}}}{1+e^{-\beta\varepsilon_{s}}}\frac{\varepsilon_{s}}{T},\label{eq:8 Thermal SNR}
\end{equation}
which is plotted in Fig. 1(b) for $M=1$. Note that for temperatures $T\ll\varepsilon_{s}$ this SNR follows a exponential scaling $\mathcal{O}\left(e^{-\frac{1}{2}\beta\varepsilon_{s}}\right)$.

\textbf{\textit{Transient regime}}. In some situations  the transient state of a system that probes a thermal bath may be more informative than its equilibrium state \cite{correa2015individual, jevtic2015single, cavina2018bridging, guo2015improved}. In our case, this occurs if the probe starts in the \textit{ground state}, see Fig. 2. The transient state $\rho_{P}^{(k)}(T)$ is completely
characterized by the ground population $p_{0,k}^{P}$, derived analytically in Sec. III of SM \cite{Note2}. Defining $r\equiv p_{1}^{s}p_{0}^{v}+p_{0}^{s}p_{1}^{v}$,
we have that 
\begin{equation}
p_{0,k}^{P}=\left[1-(1-r)^{k}\right]p_{0,\infty}^{P}+(1-r)^{k}p_{0,0}^{P},\label{eq:10 transient p0,k}
\end{equation}
where $p_{0,0}^{P}$ is the initial ground population and $p_{0,\infty}^{P}=\frac{1}{1+e^{(\beta\varepsilon_{s}-\beta_{v}\varepsilon_{v})}}$
is the steady-state value.

\begin{figure}
\includegraphics[scale=0.7]{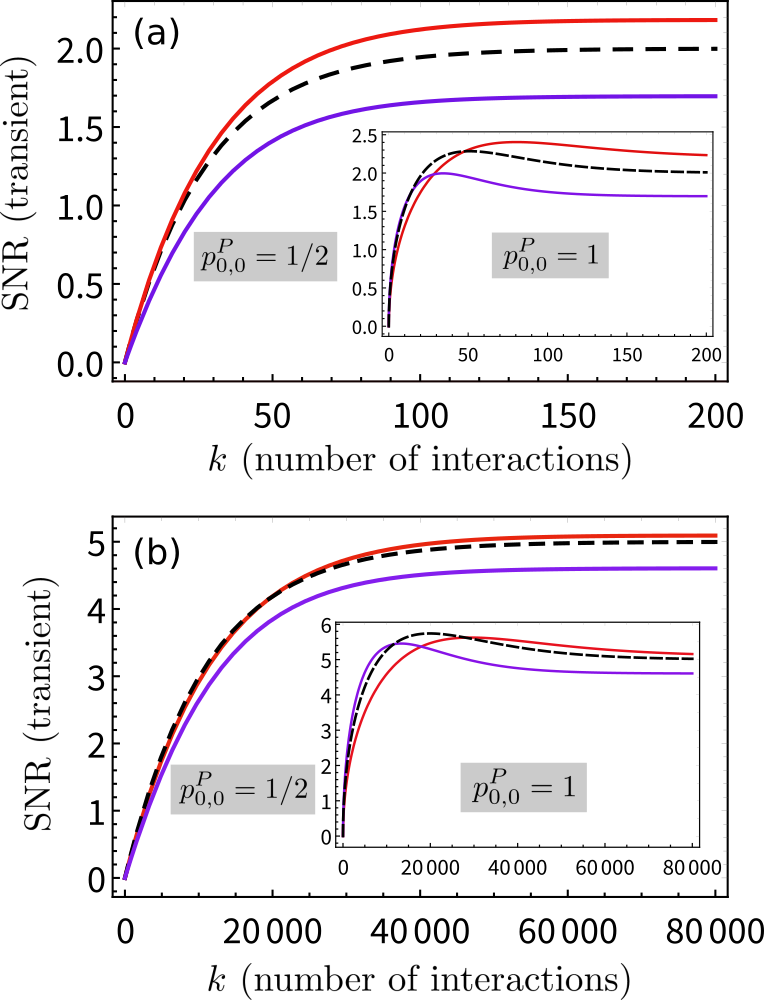}
\caption{Transient SNR (Eq. (\ref{eq:11 transient SNR})). For the main (inset) plots it is assumed that the probe starts in the fully mixed state (ground state). (a) SNR for $\widetilde{T}=\varepsilon_{s}/4$ (cf. Eq. (\ref{eq:4 Tprior})) and temperatures $T=\varepsilon_{s}/4$ (black dashed), $T=\varepsilon_{s}/4.5$ (red solid), and $T=\varepsilon_{s}/3.5$ (purple solid). (b) SNR for $\widetilde{T}=\varepsilon_{s}/10$ and temperatures $T=\varepsilon_{s}/10$ (black dashed), $T=\varepsilon_{s}/10.5$
(red solid), and $T=\varepsilon_{s}/9.5$ (purple solid). The steady-state SNR is approximately given by the asymptotic values seen in the plots.}
\end{figure}

Keeping in mind Eq. (\ref{eq:10 transient p0,k}), the transient SNR
is given by 
\begin{equation}
\frac{T}{\Delta_{k}T}=\sqrt{M}\frac{T\left|\lambda_{0,k}^{P}(T)\right|}{\sqrt{p_{0,k}^{P}(T)p_{1,k}^{P}(T)}},\label{eq:11 transient SNR}
\end{equation}
where
\begin{align}
\lambda_{0,k}^{P} & =\frac{\partial p_{0,k}^{P}}{\partial T}=\left[1-(1-r)^{k}\right]\lambda_{0,\infty}^{P}\nonumber \\
 & \quad+(k+1)\left(p_{0,\infty}^{P}-p_{0,0}^{P}\right)\frac{\partial r}{\partial T}(1-r)^{k}.\label{eq:12 transient l0}
\end{align}
The quantity $\lambda_{0,k}^{P}$ determines the sensitivity of the
population $p_{0,k}^{P}$ to small temperature variations, and is
given by $\lambda_{0,\infty}^{P}=\frac{\partial p_{0,\infty}^{P}}{\partial T}$
in the limit $k\rightarrow\infty$. 

Figure 2 shows plots of $T/\Delta_{k}T$ for $M=1$. The black dashed
curves correspond to temperatures that coincide with the prior temperature,
i.e. $T=\widetilde{T}$. The red and purple curves stand for temperatures
$T$ such that $T<\widetilde{T}$, and $T>\widetilde{T}$, respectively.
These plots show that a high thermometric precision can also be attained
in the transient regime. For $p_{0,0}^{P}=1/2$, maximum precision
is achieved for $T=\varepsilon_{s}/4.5<\widetilde{T}$ (Fig. 2(a))
and $T=\varepsilon_{s}/10.5<\widetilde{T}$ (Fig. 2(b)). If $p_{0,0}^{P}=1$
we also see that the maximum SNR takes place before reaching the steady
state, as mentioned before.

\begin{figure}
\includegraphics[scale=0.55]{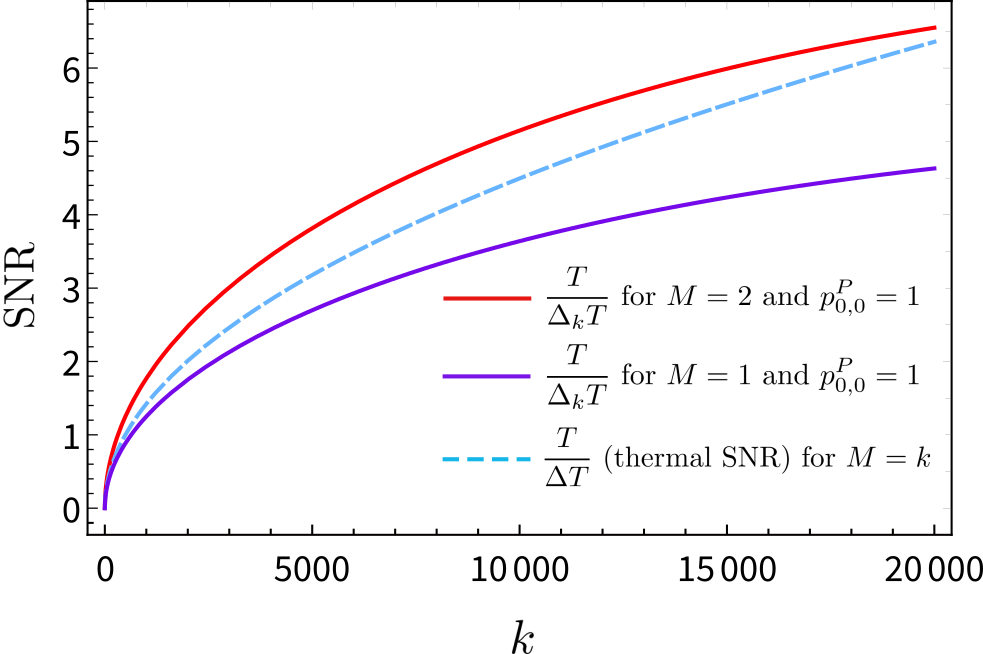}
\caption{SNR and measurement cost $M$. $T=\varepsilon_{s}/11$ and  $\widetilde{T}=\varepsilon_{s}/10$ (cf. Eq. (\ref{eq:6 ancillary tuning})) are fixed. The red and purple (solid) curves are the transient probe SNRs for $M=2$ and $M=1$, respectively, assuming that the probe starts in the ground state. The cyan (dashed) curve is the thermal SNR in Eq. (\ref{eq:8 Thermal SNR}), for $M=k$.}
\end{figure}

\textbf{\textit{Measurement cost and relation with the number of interactions}}. We define the “measurement cost” as the number of measurements $M$ required to achieve a given SNR. Figure 3 shows the thermal SNR (cyan dashed curve) and the transient SNR obtained with the machine (solid curves), for fixed parameters specified in the caption. In order to examine the corresponding measurement costs, the variable $k$ has two possible meanings. 
For the cyan dashed curve, $k=M$ is the number of measurements that must be performed on a thermalized probe to achieve the corresponding SNR. In the case of the solid curves, $k$ is the number of interactions undergone by the probe \textit{before} it is measured. The purple curve represents the SNR obtained from a single measurement ($M=1$) of the probe after $k$ interactions. The red curve is the SNR obtained from $M=2$ measurements. Since this requires to measure the same state $\rho_{P}^{(k)}(T)$ two times, the whole process must be repeated. In other words, after the first measurement the probe should be reset to its initial state and then undergo $k$ new interactions to be measured again. Remarkably, $20000$ measurements performed on the thermal probe are not enough to surpass the SNR obtained with only $2$ measurements on a probe that is prepared in the \textit{non-equilibrium state} $\rho_{P}^{(k)}(T)$. 

This analysis provides a more operational characterization of the thermometric enhancement associated with the machine, in terms of a extreme reduction of the measurement cost. Now, we discuss the "interaction cost" (i.e. the number of interactions) to achieve a given precision. After $k$ interactions, a measurement of the probe state $\rho_{P}^{(k)}(T)$ represents a suboptimal POVM performed on $k$ sample qubits. Hence, the associated SNR satisfies 
\begin{equation}
\frac{T}{\Delta_{k}T}\leq\frac{\sqrt{ke^{-\beta\varepsilon_{s}}}}{1+e^{-\beta\varepsilon_{s}}}\frac{\varepsilon_{s}}{T},\label{eq:13 inequality for number of interactions}
\end{equation}
where the r.h.s. is the optimal SNR, obtained from an energy measurement of the $k$ qubits. 

Since this bound provides the ultimate precision, we also stress that it must be obeyed by \textit{any} probing scheme. In the low-temperature regime a similar bound holds if the qubits are replaced by systems of arbitrary Hamiltonian $H_{s}=\sum_{i}\varepsilon_{i}^s|i_{s}\rangle\langle i_{s}|$, as long as the energy gap above the ground state satisfies $\varepsilon_{1}^s-\varepsilon_{0}^s\gg T$. This is a consequence of the universal scaling $\mathcal{F}_{s}^{th}(T)\sim\mathcal{O}(e^{-\beta(\varepsilon_{1}^s-\varepsilon_{0}^s)})$ that characterizes gapped systems. In particular, it implies that any sequential probing of $k$ non-interacting systems with energy gap $\varepsilon_{1}^s-\varepsilon_{0}^s\gg T$ requires at least $k\sim e^{\beta(\varepsilon_{1}^s-\varepsilon_{0}^s)}$ interactions to attain a SNR $\sim\mathcal{O}(1/T)$. 
 
\textbf{\textit{Sample perturbation}}. Reducing the back action of temperature measurements in the sample is a very desirable feature of thermometry. In Sec. VIII of \cite{Note2}, we prove that the overall heat transferred to the sample after $k$ interactions is given by
$Q_{S}^k=\varepsilon_{s}(p_{0,0}^{P}-p_{0,\infty}^{P})\left[1-(1-r)^{k}\right]$. Remarkably, we see that for $p_{0,0}^{P}>p_{0,\infty}^{P}$ the maximum heating of the sample is less than a single quantum of energy $\varepsilon_{s}$. Namely,   $\textrm{lim}_{k\rightarrow\infty}Q_{S}^k=\varepsilon_{s}\left(p_{0,0}^{P}-p_{0,\infty}^{P}\right)\leq\varepsilon_{s}$.

Apart from the negligible perturbation of the sample, our scheme allows for cooling of the sample if $p_{0,0}^{P}<p_{0,\infty}^{P}$, which is arguably a less harmful disturbance. We numerically illustrate this situation in Sec. VIII of \cite{Note2}, by considering a probe that is initialized in the fully mixed state. In contrast, for $p_{0,0}^{P}=1$ (initial ground state) the sample is inevitably heated up.            

\textbf{\textit{Conclusions}}. Heat engines harness a temperature difference for work extraction, a thermodynamic task that would otherwise be impossible. In this letter, we propose a thermometric machine that employs a high-temperature thermal bath to efficiently encode the ultracold temperature of a thermal sample in a probe. This is manifested both in the transient state of the probe and its steady state, whose maximum SNR grows as $\mathcal{O}(1/T)$ \footnote{We refer to the maximum SNR in the prior temperature interval $(0, 2 \protect\widetilde{T})$, which is achieved for $T \approx \protect\widetilde{T}$, see Fig. 1(b).}
and beats the exponential scaling that hinders low-temperature equilibrium thermometry in \textit{gapped} systems. 

The machine operates through an energy-conserving three-body interaction that characterizes absorption refrigerators \cite{levy2012quantum, correa2014quantum, correa2013performance} and other autonomous thermodynamic devices \cite{mitchison2019quantum}. In a future, it would be interesting to study if similar thermometric enhancements can be obtained using machines based on two-body interactions, which are more experimentally friendly. In Sec. IX of SM \cite{Note2}, we also show that the transient SNR (\ref{eq:11 transient SNR}) can remain above $\sim 38 \%$ of the SNR corresponding to an optimal coupling between the probe and the sample \cite{hovhannisyan2021optimal}. This result is noteworthy, keeping in mind the simplicity of the three-body interaction that drives our machine. Finally, we remark that the use of thermometric machines to probe samples comprising interacting particles is also an open subject of investigation.   
\vskip 2 mm

The authors acknowledge insightful discussions with Luis Correa. R.U. is grateful for support from Israel Science Foundation (Grant No. 2556/20).

\bibliography{references}

\setcounter{secnumdepth}{2}

\section*{Supplemental Material}

\global\long\def\thesection{S-\Roman{section}}
\setcounter{section}{0}
\global\long\def\thefigure{S\arabic{figure}}
\setcounter{figure}{0}
\global\long\def\theequation{S\arabic{equation}}
\setcounter{equation}{0}

\section{Thermometric machine for samples composed of identical, non-interacting systems}

Consider a sample formed by $N$ systems of dimension $d_{s}\geq2$,
each of which is characterized by the Hamiltonian $H_{s}=\sum_{i=0}^{d_{s}-1}\varepsilon_{s}^{(i)}|i_{s}\rangle\langle i_{s}|$.
Our goal is to show that the thermometric machine presented in the
main text can be straightforwardly adapted to this kind of sample.
More specifically, we will demonstrate that: highly precise (probe)
thermometry for $d_{s}=2$ implies highly precise thermometry for
arbitrary dimension $d_{s}$ and energy spectrum $\{\varepsilon_{s}^{(i)}\}$,
in the \textit{steady-state regime} of the probe. 

The idea is to substitute the interaction Hamiltonian $H_{I}$ in
the main text by another (three-body) Hamiltonian that couples the
probe and the ancilla to a \textit{single pair of energy levels} in
the system. Associating these energy levels with eigenstates $|j_{s}\rangle$
and $|j'_{s}\rangle$, and eigenenergies $\varepsilon_{s}^{(j')}>\varepsilon_{s}^{(j)}$,
we consider the interaction Hamiltonian
\begin{equation}
H_{I}=|0_{P}j_{s}1_{v}\rangle\langle1_{P}j'_{s}0_{v}|+\textrm{h.c.}\label{eq:12.1 general interaction}
\end{equation}
In this way, we will show that the repeated switching of $H_{I}$
drives the probe through ``probabilistic collisions'' with the ancilla
and an effective qubit of energy gap $\varepsilon_{s}^{(j')}-\varepsilon_{s}^{(j)}$,
equilibrated at the temperature of the sample. This dynamics leads
the probe to a steady state that allows us to translate the results
obtained for dimension $d_{s}=2$ to arbitrary dimension. 

For $d_{s}$ arbitrary, the total free Hamiltonian $H_{\textrm{free}}=H_{P}+H_{s}+H_{v}$
has eigenstates $\{|0_{P}\rangle,|1_{P}\rangle\}\otimes\{|i_{s}\rangle\}_{i=0}^{d_{s}-1}\otimes\{|0_{v}\rangle,|1_{v}\rangle\}$.
Since $H_{I}$ transforms eigenstates outside the subspace $\textrm{span}[\{|0_{P}\rangle,|1_{P}\rangle\}\otimes\{|j_{s}\rangle,|j'_{s}\rangle\}\otimes\{|0_{v}\rangle,|1_{v}\rangle\}]$
to the zero vector, these eigenstates are also eigenstates of the
total Hamiltonian $H=H_{\textrm{free}}+H_{I}$. Based on this property,
we can conveniently express the evolution generated by $H$ on an
initial state of the form $\rho=\rho_{P}\otimes\rho_{s}\otimes\rho_{v}$,
where $\rho_{s}$ is a thermal state and $\rho_{P}$ and $\rho_{v}$
are both diagonal in the corresponding energy bases. To this end,
we write $\rho_{s}$ as $\rho_{s}=\sum_{i=0}^{d_{s}-1}p_{i}^{s}(T)|i_{s}\rangle\langle i_{s}|$,
where $p_{i}^{s}(T)=\frac{e^{-\beta\varepsilon_{s}^{(i)}}}{\Tr\left(e^{-\beta H_{s}}\right)}$
are thermal populations at temperature $T$, and the total state as
\begin{equation}
\rho=(1-p_{j}^{s}(T)-p_{j'}^{s}(T))\sigma+(p_{j}^{s}(T)+p_{j'}^{s}(T))\rho_{j,j'},\label{eq:12.2 decomposition of rho}
\end{equation}
where 

\begin{align}
\sigma&=\frac{\rho_{P}\otimes\rho_{v}\otimes\sum_{i\neq j,j'}p_{i}^{s}(T)|i_{s}\rangle\langle i_{s}|}{(1-p_{j}^{s}(T)-p_{j'}^{s}(T))},\nonumber
\\
\rho_{j,j'}&
=\rho_{P}\otimes\rho_{v}\otimes\rho_{s;j,j'},\nonumber
\\
\rho_{s;j,j'}&=\frac{p_{j}^{s}(T)|j_{s}\rangle\langle j_{s}|+p_{j'}^{s}(T)|j'_{s}\rangle\langle j'_{s}|}{p_{j}^{s}(T)+p_{j'}^{s}(T)}.\label{eq:12.3 rho(j,j')}
\end{align}

As anticipated
at the beginning of this appendix, we see that $\rho_{s;j,j'}$
is the thermal state (at temperature $T$) of an effective two-level
level system characterized by the Hamiltonian $H_{s}^{\textrm{eff}}=\varepsilon_{s}^{(j)}|j_{s}\rangle\langle j_{s}|+\varepsilon_{s}^{(j')}|j'_{s}\rangle\langle j'_{s}|$. Defining the map 
\begin{equation}
	\mathcal{E}_{P:j,j'}(\rho_{P})\equiv\Tr_{sv}\left[e^{-iHt}\rho_{P}\otimes\rho_{v}\otimes\rho_{s;j,j'}e^{iHt}\right],\label{eq:12.6 map Ej,j'}
\end{equation}
it follows that a collision with the probe is described by the map 
\begin{align}
	\mathcal{E}_{P}(\rho_{P})&\equiv\Tr_{sv}\left[e^{-iHt}\rho_{P}\otimes\rho_{v}\otimes\rho_{s} e^{iHt}\right]\nonumber \\ & =(1-p_{j}^{s}(T)-p_{j'}^{s}(T))e^{-iH_{P}t}\rho_{P} e^{iH_{P}t}\nonumber \\
	& \quad+(p_{j}^{s}(T)+p_{j'}^{s}(T))\mathcal{E}_{P:j,j'}(\rho_{P}),\label{eq:12.5 evoution on rho}
\end{align}
which characterizes a transformation where a collision with the aforementioned two-level system occurs with probability \textit{$p_{j}^{s}(T)+p_{j'}^{s}(T)$}.
    
In contrast, the maps $	\mathcal{E}_{P}$  and $	\mathcal{E}_{P;j,j'}$ posses a common (and unique) fixed point $\rho_{P;j,j'}^{(\infty)}$, which implies that collisions with the effective two-level system fully determine the  steady state of the probe. This follows from the fact that  \textit{$\rho_{P;j,j'}^{(\infty)}$ is diagonal in the energy basis of the probe}, as shown in the next appendix. In such a case, a direct substitution of $\rho_{P}$ by  $\rho_{P;j,j'}^{(\infty)}$ in Eq. (\ref{eq:12.5 evoution on rho}) shows that $\mathcal{E}_{P}(\rho_{P;j,j'}^{(\infty)})=\mathcal{E}_{P;j,j'}(\rho_{P;j,j'}^{(\infty)})$. Note  also that this is possible because  $e^{-iH_{P}t}\rho_{P} e^{iH_{P}t}=\rho_{P}$ for any $\rho_{P}$ diagonal. 

In the next appendix we derive the probe evolution corresponding to sample subsystems of dimension $d_s=2$, for which we find that the steady state is diagonal in the eigenbasis of $H_{P}$. Since this evolution amounts to repeated applications of the map $\mathcal{E}_{P;j,j'}$, the diagonal character of  $\rho_{P;j,j'}$ follows. In this way we prove the main claim of this appendix. Namely, that for the interaction (\ref{eq:12.1 general interaction}) the corresponding steady state can be described in terms of collisions with the ancilla and thermal qubits.

\section{Derivation of the global (probe-sample-ancilla) evolution}

In this appendix we solve the Schrodinger equation for the total Hamiltonian
$H=H_{P}+H_{s}+H_{v}+H_{I}$. If $\varepsilon_{v}=\varepsilon_{s}+\varepsilon_{P}$,
the interaction Hamitonian $H_{I}=|0_{P}0_{s}1_{v}\rangle\langle1_{P}1_{s}0_{v}|+\textrm{h.c.}$
commutes with the free Hamiltonian $H_{\textrm{free}}=H_{P}+H_{s}+H_{v}$.
This follows from the commutation relations 
\begin{align}
\left[H_{s},H_{I}\right] & =\varepsilon_{s}\varepsilon_{I}\left(-|0_{P}0_{s}1_{v}\rangle\langle1_{P}1_{s}0_{v}|+\textrm{h.c.}\right),\label{eq:10}\\
\left[H_{P},H_{I}\right] & =\varepsilon_{P}\varepsilon_{I}\left(-|0_{P}0_{s}1_{v}\rangle\langle1_{P}1_{s}0_{v}|+\textrm{h.c.}\right),\label{eq:11}\\
\left[H_{v},H_{I}\right] & =\varepsilon_{v}\varepsilon_{I}\left(|0_{P}0_{s}1_{v}\rangle\langle1_{P}1_{s}0_{v}|-\textrm{h.c.}\right),\label{eq:12}
\end{align}
which, when added together, yield a commutator $\left[H_{\textrm{free}},H_{I}\right]$
proportional to $\varepsilon_{v}-\varepsilon_{s}-\varepsilon_{P}$.
Therefore, the evolution during a time $t$ is given by the global
unitary 
\begin{equation}
U_{Psv}(t)=e^{-iHt}=e^{-iH_{\textrm{free}}t}e^{-iH_{I}t}.\label{eq:13 global unitary}
\end{equation}

The eigenstates of $U_{Psv}(t)$ can be obtained by diagonalizing
$H_{I}$. In the eigenbasis of $H_{\textrm{free}}$, $H_{I}$ is a
matrix with all the entries equal to zero except for $\langle0_{P}0_{s}1_{v}|H_{I}|1_{P}1_{s}0_{v}\rangle=\langle1_{P}1_{s}0_{v}|H_{I}|0_{P}0_{s}1_{v}\rangle=\varepsilon_{I}$.
The straightforward diagonalization leads to 
\begin{align}
H_{I}|+\rangle & =\varepsilon_{I}|+\rangle,\label{eq:14 HI on |+>}\\
H_{I}|-\rangle & =-\varepsilon_{I}|-\rangle,\label{eq:15 HI on |->}
\end{align}
where $|\pm\rangle\equiv\frac{1}{\sqrt{2}}\left(|0_{P}0_{s}1_{v}\rangle\pm|1_{P}1_{s}0_{v}\rangle\right)$. 

Expressing $|0_{P}0_{s}1_{v}\rangle$ as $|0_{P}0_{s}1_{v}\rangle=\frac{1}{\sqrt{2}}\left(|+\rangle+|-\rangle\right)$
and $|1_{P}1_{s}0_{v}\rangle$ as $|1_{P}1_{s}0_{v}\rangle=\frac{1}{\sqrt{2}}\left(|+\rangle-|-\rangle\right)$,
we have that 
\begin{align}
e^{-iH_{I}t}|0_{P}0_{s}1_{v}\rangle & =\frac{1}{\sqrt{2}}\left(e^{-i\varepsilon_{I}t}|+\rangle+e^{i\varepsilon_{I}t}|-\rangle\right),\label{eq:16 e(-HI) on 001}\\
e^{-iH_{I}t}|1_{P}1_{s}0_{v}\rangle & =\frac{1}{\sqrt{2}}\left(e^{-i\varepsilon_{I}t}|+\rangle-e^{i\varepsilon_{I}t}|-\rangle\right).\label{eq:17 e(-HI) on 110}
\end{align}
Accordingly, for $t=\frac{\pi}{2\varepsilon_{I}}$, $e^{-iH_{I}t}|0_{P}0_{s}1_{v}\rangle=-i|1_{P}1_{s}0_{v}\rangle$
and $e^{-iH_{I}t}|1_{P}1_{s}0_{v}\rangle=-i|0_{P}0_{s}1_{v}\rangle$.
Since $|0_{P}0_{s}1_{v}\rangle$ and $|1_{P}1_{s}0_{v}\rangle$ are
both eigenstates of $H_{\textrm{free}}$, Eq. (\ref{eq:13 global unitary})
yields
\begin{align}
U_{Psv}(\pi/2\varepsilon_{I})|0_{P}0_{s}1_{v}\rangle & \propto|1_{P}1_{s}0_{v}\rangle,\label{eq:18 UPsv on 001}\\
U_{Psv}(\pi/2\varepsilon_{I})|1_{P}1_{s}0_{v}\rangle & \propto|0_{P}0_{s}1_{v}\rangle.\label{eq:19 UPsv on 110}
\end{align}
On the other hand, $H_{I}|i_{P}j_{s}k_{v}\rangle=0$ for any eigenstate
of $H_{\textrm{free}}$ different from $|0_{P}0_{s}1_{v}\rangle$
and $|1_{P}1_{s}0_{v}\rangle$, and thus 
\begin{equation}
U_{Psv}(\pi/2\varepsilon_{I})|i_{P}j_{s}k_{v}\rangle\propto|i_{P}j_{s}k_{v}\rangle.\label{eq:20 UPsv(pi/2)}
\end{equation}

If $U_{Psv}(\pi/2\varepsilon_{I})$ is applied on any state diagonal
in the eigenbasis of $H_{\textrm{free}}$, the phase factors implicit
in Eqs. (\ref{eq:18 UPsv on 001})-(\ref{eq:20 UPsv(pi/2)}) cancel
out. This is the case in particular for the initial state $\rho_{P}^{(0)}\otimes\rho_{s}(T)\otimes\rho_{v}$,
where $\rho_{P}^{(0)}$, $\rho_{s}(T)$ and $\rho_{v}$ are diagonal
in the corresponding energy bases. Moreover, any application of $U_{Psv}(\pi/2\varepsilon_{I})$
yields another diagonal state. This implies that for any state $\rho_{P}^{(k)}\otimes\rho_{s}(T)\otimes\rho_{v}$
(where $\rho_{P}^{(k)}$ is the probe state after $k$ cycles, see
main text), the aforementioned phases can be ignored in the subsequent
evolution generated by $U_{Psv}(\pi/2\varepsilon_{I})$. 

\section{Probe state after $k$ interactions }

As mentioned above, the state $\rho_{P}^{(k)}$ is diagonal in the
eigenbasis of $H_{P}$. Hence, it suffices to compute the population
$p_{0,k}^{P}$ to fully characterize this state. For the sake of brevity,
in the following we will write $U_{Psv}(\pi/2\varepsilon_{I})$ as
$U_{Psv}$. The state $\rho_{P}^{(k+1)}$ is is related to the previous
state $\rho_{P}^{(k)}$ through the expression:
\begin{equation}
\rho_{P}^{(k+1)}=\Tr_{sv}\left[U_{Psv}\left(\rho_{P}^{(k)}\otimes\rho_{s}\otimes\rho_{v}\right)U_{Psv}^{\dagger}\right].\label{eq:21 rho(k+1) in terms of rho(k)}
\end{equation}

From Eqs.~(\ref{eq:18 UPsv on 001})-(\ref{eq:20 UPsv(pi/2)}), $U_{Psv}$
exchanges the populations of the eigenstates $|0_{P}0_{s}1_{v}\rangle$
and $|1_{P}1_{s}0_{v}\rangle$, and keeps any other population unmodified.
Therefore, 
\begin{equation}
p_{0,k+1}^{P}=p_{0,k}^{P}+\Delta_{k+1}p,\label{eq:22 p0,k+1 in terms of p0,k}
\end{equation}
where $\Delta_{k+1}p=p_{0,k+1}^{P}-p_{0,k}^{P}$ is the population
variation of the eigenstate $|0_{P}0_{s}1_{v}\rangle$, resulting
from the $(k+1)$th cycle. Explicitly, 
\begin{align}
\Delta_{k+1}p & =p_{1,k}^{P}\left(p_{1}^{s}p_{0}^{v}\right)-p_{0,k}^{P}\left(p_{0}^{s}p_{1}^{v}\right)\nonumber \\
 & =p_{1}^{s}p_{0}^{v}-rp_{0,k}^{P},\label{eq:23 delta(k+1)p}
\end{align}
where 
\begin{equation}
r\equiv p_{1}^{s}p_{0}^{v}+p_{0}^{s}p_{1}^{v}\label{eq:24 definition of r}
\end{equation}
and the second line follows by writing $p_{1,k}^{P}=1-p_{0,k}^{P}$. 

In the limit $k\rightarrow\infty$, the population $p_{0,k}^{P}$
reaches the steady-state value and thus $\Delta_{k+1}p=0$. Hence,
from Eq. (\ref{eq:23 delta(k+1)p}) we obtain:
\begin{equation}
p_{0,\infty}^{P}=\frac{p_{1}^{s}p_{0}^{v}}{r}=\frac{1}{1+e^{\beta\varepsilon_{s}-\beta_{v}\varepsilon_{v}}}.\label{eq:25 p0,inf}
\end{equation}
Using this expression we can also rewrite Eq. (\ref{eq:23 delta(k+1)p})
as $\Delta_{k+1}p=r\left(p_{0,\infty}^{P}-p_{0,k}^{P}\right)$, and
Eq. (\ref{eq:22 p0,k+1 in terms of p0,k}) as 
\begin{equation}
p_{0,k+1}^{P}=(1-r)p_{0,k}^{P}+rp_{0,\infty}^{P}.\label{eq:26 p0,k+1 recurrence relation}
\end{equation}

Let us see now that (for any $k\geq0$) the recurrence relation (\ref{eq:26 p0,k+1 recurrence relation})
is satisfied by the expression 
\begin{align}
p_{0,k+1}^{P} & =\sum_{j=0}^{k}(1-r)^{j}rp_{0,\infty}^{P}+(1-r)^{k+1}p_{0,0}^{P}\nonumber \\
 & =\left[1-(1-r)^{k+1}\right]p_{0,\infty}^{P}+(1-r)^{k+1}p_{0,0}^{P},\label{eq:27 p0,k+1 explicit expression}
\end{align}
where the second line is obtained by computing the geometric series
$\sum_{j=0}^{k}(1-r)^{j}$. This can be straightforwardly proved by
induction. For $k=0$, Eq. (\ref{eq:27 p0,k+1 explicit expression})
yields $p_{0,1}^{P}=rp_{0,\infty}^{P}+(1-r)p_{0,0}^{P}$, in agreement
with Eq. (\ref{eq:26 p0,k+1 recurrence relation}). Moreover, the
substitution of $p_{0,k+1}^{P}$ (as given in Eq. (\ref{eq:27 p0,k+1 explicit expression}))
into $p_{0,k+2}^{P}=(1-r)p_{0,k+1}^{P}+rp_{0,\infty}^{P}$ (cf. (\ref{eq:26 p0,k+1 recurrence relation}))
yields 
\begin{align}
p_{0,k+2}^{P} & =(1-r)\left[1-(1-r)^{k+1}\right]p_{0,\infty}^{P}+(1-r)^{k+2}p_{0,0}^{P}\nonumber \\
 & \quad+rp_{0,\infty}^{P}\nonumber \\
 & =\left[1-(1-r)^{k+2}\right]p_{0,\infty}^{P}+(1-r)^{k+2}p_{0,0}^{P}.\label{eq:28 verification for k+1}
\end{align}
This means that if Eq. (\ref{eq:27 p0,k+1 explicit expression}) holds
for $k$, then it also holds for $k+1$. Since it is valid for $k=0$,
it also follows that it is valid for any $k\geq0$. 

\section{Signal-to-Noise Ratio (SNR) for energy measurements on $\rho_{P}^{(k)}$}

Here we derive the thermometric precision corresponding to energy
measurements on the probe state $\rho_{P}^{(k)}(T)$. As a byproduct,
we show also that these measurements saturate the associated Cram\'{e}r-Rao
bound. If $\Delta_{k}T$ denotes the (absolute) error for the estimation
of $T$ on $\rho_{P}^{(k)}(T)$, the aforementioned saturation means
that 
\begin{equation}
\Delta_{k}T=\frac{1}{\sqrt{M\mathcal{F}_{P}^{(k)}(T)}},\label{eq:29 absolute error after k interactions}
\end{equation}
where 
\begin{align}
\mathcal{F}_{P}^{(k)}(T) & =\sum_{i=0}^{1}p_{i,k}^{P}(T)\left(\frac{\partial\textrm{ln}(p_{i,k}^{P})}{\partial T}\right)^{2}\nonumber \\
 & =\frac{1}{p_{0,k}^{P}(T)p_{1,k}^{P}(T)}\left(\frac{\partial p_{1,k}^{P}}{\partial T}\right)^{2}\label{eq:30 FI after k interactions}
\end{align}
is the Fisher information that results from energy measurements on
$\rho_{P}^{(k)}(T)$. 

Consider the error propagation formula \cite{mehboudi2019thermometry}
\begin{equation}
\Delta T=\frac{\sqrt{\Var(O)}}{\sqrt{M}\left|\frac{\partial\bigl\langle O\bigr\rangle}{\partial T}\right|},\label{eq:31 error propagation formula}
\end{equation}
which provides an expression for $\Delta T$, given that $T$ is estimated
from $M$ (independent) measurements of an observable $O$. In this
formula, $\Var(O)=\Tr\left[O^{2}\rho(T)\right]-\left(\Tr\left[O\rho(T)\right]\right)^{2}$
and $\bigl\langle O\bigr\rangle=\Tr\left[O\rho(T)\right]$
are respectively the variance and mean value of $O$, with respect
to a state $\rho(T)$ where the temperature $T$ has been encoded.
If $O=H_{P}=\varepsilon_{P}|1_{P}\rangle\langle1_{P}|$ and $\rho(T)=\rho_{P}^{(k)}(T)$,
we have that 
\begin{align}
\Var\left(H_{P}\right) & =p_{0,k}^{P}(T)p_{1,k}^{P}(T)\varepsilon_{P}^{2},\label{eq:32 variance after k interactions}\\
\frac{\partial\bigl\langle H_{P}\bigr\rangle}{\partial T} & =\frac{\partial p_{1,k}^{P}(T)}{\partial T}\varepsilon_{P}\nonumber \\
 & =-\frac{\partial p_{0,k}^{P}(T)}{\partial T}\varepsilon_{P},\label{eq:33 derivative of <HP> after k interactions}
\end{align}
where the third line follows from probability conservation: $\frac{\partial\left(p_{0,k}^{P}(T)+p_{0,k}^{P}(T)\right)}{\partial T}=0$. 

Let us now define the ``sensitivity'' 
\begin{equation}
\lambda_{0,k}^{P}(T)\equiv\frac{\partial p_{0,k}^{P}(T)}{\partial T}.\label{eq:33.1 sensitivity}
\end{equation}
In this way, the substitution of Eqs. (\ref{eq:32 variance after k interactions})
and (\ref{eq:33 derivative of <HP> after k interactions})
into Eq. (\ref{eq:31 error propagation formula}) lead to 
\begin{equation}
\Delta_{k}T=\frac{\sqrt{p_{0,k}^{P}(T)p_{1,k}^{P}(T)}}{\sqrt{M}\left|\lambda_{0,k}^{P}(T)\right|}.\label{eq:34 absolute error explicit}
\end{equation}
By comparing Eqs. (\ref{eq:30 FI after k interactions}) and (\ref{eq:34 absolute error explicit}),
we see that the saturation of the Cram\'{e}r-Rao bound (cf. Eq. (\ref{eq:29 absolute error after k interactions}))
follows.

\section{Steady-state SNR (limits $N\rightarrow\infty$ and $k\rightarrow\infty$) }

The steady state SNR is obtained by taking the limit $\textrm{lim}_{k\rightarrow\infty}\Delta_{k}T$
(cf. Eq. (\ref{eq:34 absolute error explicit})), in the expression
$T/\Delta_{k}T$. From Eqs. (\ref{eq:33.1 sensitivity}) and (\ref{eq:25 p0,inf}),
it follows that 
\begin{align}
\lambda_{0,\infty}^{P}(T) & =\frac{\partial p_{0,\infty}^{P}}{\partial T}\nonumber \\
 & =\frac{e^{\beta\varepsilon_{s}-\beta_{v}\varepsilon_{v}}}{(1+e^{\beta\varepsilon_{s}-\beta_{v}\varepsilon_{v}})^{2}}\frac{\varepsilon_{s}}{T^{2}}\nonumber \\
 & =p_{0,\infty}^{P}(T)p_{1,\infty}^{P}(T)\frac{\varepsilon_{s}}{T^{2}}.\label{eq:35 steady-state sensitivity}
\end{align}
Therefore, 
\begin{align}
\frac{T}{\Delta_{\infty}T} & =\sqrt{M\left(p_{0,\infty}^{P}(T)p_{1,\infty}^{P}(T)\right)}\frac{\varepsilon_{s}}{T}\nonumber \\
 & =\frac{\sqrt{Me^{\beta\varepsilon_{s}-\beta_{v}\varepsilon_{v}}}}{1+e^{\beta\varepsilon_{s}-\beta_{v}\varepsilon_{v}}}\frac{\varepsilon_{s}}{T}\nonumber \\
 & =\sqrt{M}\frac{e^{-\frac{1}{2}(\beta\varepsilon_{s}-\beta_{v}\varepsilon_{v})}}{1+e^{-(\beta\varepsilon_{s}-\beta_{v}\varepsilon_{v})}}\frac{\varepsilon_{s}}{T}.\label{eq:36 steady-state SNR}
\end{align}

\section{Steady-state SNR for imprecise ancillary temperature}

Previously we have computed the SNR assuming that the temperature
$T_{v}$ is perfectly known. However, in practice this temperature
must also be estimated in order to evaluate the ancillary energy gap
$\varepsilon_{v}$ (cf. (\ref{eq:6 ancillary tuning}) in the main
text), and such an estimation carries a finite error $\Delta T_{v}$.
Denoting the estimated ancillary temperature as $T_{v}^{\textrm{est}}$and
the actual value as $T_{v}$, we thus have that $T_{v}^{\textrm{est}}=T_{v}\pm\Delta T_{v}$.
The purpose of this appendix is to show that the steady-state SNR
for $T$ is quite robust to the error $\Delta T_{v}$, if the corresponding
SNR satisfies $T_{v}/\Delta T_{v}\geq2$. Since $T_{v}$ can be larger
than $T$, the exponential inefficiency in the estimation of $T$
does not carry over the estimation of $T_{v}$. For example, one could
estimate $T_{v}$ via measurements of the Hamiltonian $H_{u}=\varepsilon_{u}|1_{u}\rangle\langle1_{u}|$
of a two-level system that has been thermalized at temperature $T_{v}$
and whose energy gap is such that $T_{v}/\Delta T_{v}$ is maximized.
Specifically, the thermal SNR corresponding to $M$ energy measurements
on $\rho_{u}=\frac{e^{-\beta_{v}H_{u}}}{\Tr(e^{-\beta_{v}H_{u}})}$
reads 
\begin{equation}
\frac{T_{v}}{\Delta T_{v}}=\sqrt{M}\frac{e^{-\frac{1}{2}\beta_{v}\varepsilon_{u}}}{1+e^{-\beta_{v}\varepsilon_{u}}}\frac{\varepsilon_{u}}{T_{v}}.\label{eq:36.1 thermal SNR for Tv}
\end{equation}
This quantity attains its maximum $\textrm{max}_{\varepsilon_{u}}T_{v}/\Delta T_{v}\sim\sqrt{M}0.66$
for $\beta_{v}\varepsilon_{u}\sim2.5$. Hence, $M=16$ measurements
suffice to yield an SNR $T_{v}/\Delta T_{v}\sim2.6>2$. 

To analyze the impact of the error $\Delta T_{v}$ on the SNR that
characterizes the estimation of $T$ it is convenient to recall the
error-free expression (Eq. (\ref{eq:5 steady SNR}) in the main text)
\begin{equation}
\frac{T}{\Delta_{\infty}T}=\frac{\sqrt{M}e^{-\frac{1}{2}\beta\varepsilon_{s}\left(1-\frac{T}{\widetilde{T}}\right)}}{1+e^{-\beta\varepsilon_{s}\left(1-\frac{T}{\widetilde{T}}\right)}}\frac{\varepsilon_{s}}{T},\label{eq:36.1 Steady-S SNR in terms of Tprior}
\end{equation}
where we have written $T_{\textrm{max}}$ as $2\widetilde{T}$ ((\ref{eq:4 Tprior})
in the main text). We note that for $0<T<2\widetilde{T}$, which is
the interval where the temperature $T$ is known to be, the factor
$1-\frac{T}{\widetilde{T}}$ varies in the interval $(-1,1)$. Hence,
for the extremal points -1 and 1 of this interval we have that $\frac{e^{-\frac{1}{2}\beta\varepsilon_{s}\left(1-\frac{T}{\widetilde{T}}\right)}}{1+e^{-\beta\varepsilon_{s}\left(1-\frac{T}{\widetilde{T}}\right)}}=\frac{e^{-\frac{1}{2}\beta\varepsilon_{s}}}{1+e^{-\beta\varepsilon_{s}}}$
and consequently Eq. (\ref{eq:36.1 Steady-S SNR in terms of Tprior})
yields the thermal SNR corresponding to temperatures $T\rightarrow0$
and $T=2\widetilde{T}$. On the other hand, for all $T\in(0,2\widetilde{T})$
the ratio $\frac{e^{-\frac{1}{2}\beta\varepsilon_{s}\left(1-\frac{T}{\widetilde{T}}\right)}}{1+e^{-\beta\varepsilon_{s}\left(1-\frac{T}{\widetilde{T}}\right)}}$
is \textit{strictly large}r than $\frac{e^{-\frac{1}{2}\beta\varepsilon_{s}}}{1+e^{-\beta\varepsilon_{s}}}$,
attaining a maximum of $1/2$ if $T=\widetilde{T}$. This occurs because
the factor $1-\frac{T}{\widetilde{T}}$ ``damps'' the exponent $\beta\varepsilon_{s}$
and prevents that the exponential $e^{-\frac{1}{2}\beta\varepsilon_{s}\left(1-\frac{T}{\widetilde{T}}\right)}$
vanishes even if $\beta\varepsilon_{s}\gg1$. A similar situation
occurs if $\Delta T_{v}>0$ and $T_{v}/\Delta T_{v}\geq2$, as we
discuss below. 

If $\Delta T_{v}>0$ the SNR is modified via the error in the tuning
of the energy gap $\varepsilon_{v}$, which is chosen according to
$T_{v}^{\textrm{est}}$ instead of $T_{v}$: 
\begin{equation}
\varepsilon_{v}=\frac{\varepsilon_{s}}{\widetilde{T}}T_{v}^{\textrm{est}}=\frac{\varepsilon_{s}}{\widetilde{T}}(T_{v}\pm\Delta T_{v}).\label{eq:36.2 noisy tuning of energy gap}
\end{equation}
In this way, instead of the ideal expression (\ref{eq:36.1 Steady-S SNR in terms of Tprior})
the steady-state SNR $\frac{\sqrt{M}e^{-\frac{1}{2}(\beta\varepsilon_{s}-\beta_{v}\varepsilon_{v})}}{1+e^{-(\beta\varepsilon_{s}-\beta_{v}\varepsilon_{v})}}\frac{\varepsilon_{s}}{T}$
(cf. Eq. (\ref{eq:36 steady-state SNR})) takes the form 
\begin{align}
\frac{T}{\Delta_{\infty}T} & =\frac{\sqrt{M}e^{-\frac{1}{2}\beta\varepsilon_{s}\left[1-\frac{T}{\widetilde{T}}\left(1\pm\frac{\Delta T_{v}}{T_{v}}\right)\right]}}{1+e^{-\beta\varepsilon_{s}\left[1-\frac{T}{\widetilde{T}}\left(1\pm\frac{\Delta T_{v}}{T_{v}}\right)\right]}}\frac{\varepsilon_{s}}{T}\nonumber \\
 & \equiv\frac{\sqrt{M}e^{-\frac{1}{2}\beta\varepsilon_{s}x_{T}}}{1+e^{-\beta\varepsilon_{s}x_{T}}}\frac{\varepsilon_{s}}{T}.\label{eq:36.3 noisy steady-state SNR}
\end{align}
Here, the factor $x_{T}\equiv1-\frac{T}{\widetilde{T}}\left(1\pm\frac{\Delta T_{v}}{T_{v}}\right)$
is what determines the damping of the exponent $\beta\varepsilon_{s}$,
and as long as $e^{-\frac{1}{2}\beta\varepsilon_{s}x_{T}}$ remains
finite in the limit $\beta\varepsilon_{s}\gg1$ we can guarantee that
the non-ideal SNR (\ref{eq:36.3 noisy steady-state SNR}) is not exponentially
suppressed. In particular, for $x_{T}=0$ we have that 
\begin{equation}
T=\frac{\tilde{T}}{1\pm\frac{\Delta T_{v}}{T_{v}}},\label{eq:36.4 temperatures that maximize the non-ideal SNR}
\end{equation}
and 
\begin{equation}
\frac{T}{\Delta_{\infty}T}=\frac{\sqrt{M}}{2}\left(1\pm\frac{\Delta T_{v}}{T_{v}}\right)\frac{\varepsilon_{s}}{\tilde{T}}.\label{eq:36.5 "Maximum" noisy SNR}
\end{equation}

This expression indicates that the error $\Delta T_{v}$ does not
alter the scaling $\mathcal{O}(\varepsilon_{s}/\tilde{T})$ previously
found for the ideal SNR (as the only difference is the constant $1\pm\frac{\Delta T_{v}}{T_{v}}$).
However, it is worth noting that if $T=\frac{\tilde{T}}{1-\frac{\Delta T_{v}}{T_{v}}}$
then $T\in(0,2\tilde{T})$ if and only if $T_{v}/\Delta T_{v}\geq2$.
Keeping in mind that the plus and minus signs in Eqs. (\ref{eq:36.4 temperatures that maximize the non-ideal SNR})
and (\ref{eq:36.5 "Maximum" noisy SNR}) correspond respectively to
estimated temperatures $T_{v}+\Delta T_{v}$ and $T_{v}-\Delta T_{v}$,
the condition $T_{v}/\Delta T_{v}\geq2$ ensures that if $T_{v}^{\textrm{est}}=T_{v}-\Delta T_{v}$
the SNR $\frac{\sqrt{M}}{2}\left(1-\frac{\Delta T_{v}}{T_{v}}\right)\frac{\varepsilon_{s}}{\tilde{T}}$
can be obtained for some temperature in the prior interval $(0,2\tilde{T})$.
Finally, we note that the SNR (\ref{eq:36.3 noisy steady-state SNR})
also decays to exponentially small values as $T$ approaches the extrema
0 and $2\widetilde{T}$. However, this is not different from the ideal
SNR (corresponding to $\Delta T_{v}=0$), as discussed before. 

\section{Transient SNR }

The transient SNR is obtained by substituting Eq. (\ref{eq:27 p0,k+1 explicit expression})
into (\ref{eq:34 absolute error explicit}) and evaluating explicitly
$\lambda_{0,k}^{P}(T)$. To compute $\lambda_{0,k}^{P}(T)$ we only
have to take the derivative of (\ref{eq:34 absolute error explicit})
with respect to $T$: 
\begin{align}
\lambda_{0,k+1}^{P} & =\frac{\partial p_{0,k+1}^{P}}{\partial T}\nonumber \\
 & =\left[1-(1-r)^{k+1}\right]\lambda_{0,\infty}^{P}\nonumber \\
 & \quad+(k+1)\left(p_{0,\infty}^{P}-p_{0,0}^{P}\right)\frac{\partial r}{\partial T}(1-r)^{k}.\label{eq:37 transient sensitivity}
\end{align}
Here, $\lambda_{0,\infty}^{P}=\frac{\partial p_{0,\infty}^{P}}{\partial T}$
is obtained from Eq. (\ref{eq:25 p0,inf}) and $\frac{\partial r}{\partial T}$
is obtained from Eq. (\ref{eq:24 definition of r}). 

\section{Perturbation of the sample }

In this appendix we present analytical and numerical results concerning the perturbation
of the sample and ancillary states during the transient dynamics.
These results indicate that preparing the probe in a completely mixed
state generates less perturbation, as compared to preparing it in
the ground state. Moreover, the sample is always heated up in the
second case, while it can some times be cooled down if the probe starts
in the fully mixed state. 

We first apply Eq. (\ref{eq:27 p0,k+1 explicit expression}) to derive the total heat transferred to the sample and to the ancillary bath after $k$ interactions. Due to the form of $H_{I}$, at the $(k+1)$th cycle two situations can take place. If the ground state of the probe gains population $\Delta_{k+1}p=p_{0,k+1}^P-p_{0,k}^P$, the corresponding sample qubit releases heat $\Delta_{k+1}p\varepsilon_{s}$ and the ancilla absorbs heat $\Delta_{k+1}p\varepsilon_{v}$, which equals the heat absorbed by the ancillary bath in the subsequent thermalization. The opposite situation occurs if $\Delta_{k+1}p<0$. 

After $k$ cycles, the sample and the ancilla have absorbed heat $Q_{S}^k$ and $Q_{v}^k$, respectively. The heat intake to the sample corresponding to the $j$th cycle reads $-\Delta_{j+1}p\varepsilon_{s}$. Moreover, the ancilla absorbs heat $\Delta_{j+1}p\varepsilon_{v}$. Using Eq. (\ref{eq:27 p0,k+1 explicit expression}), we obtain the total heats  
\begin{align}
Q_{S}^k&=-\sum_{j=0}^{k-1}\Delta_{j+1}p\varepsilon_{s}=\varepsilon_{s}\left(p_{0,0}^{P}-p_{0,\infty}^{P}\right)\left[1-(1-r)^{k}\right],\label{eq:38 sample heat}\\
Q_{v}^k&=\sum_{j=0}^{k-1}\Delta_{j+1}p\varepsilon_{v}=\varepsilon_{v}\left(p_{0,\infty}^{P}-p_{0,0}^{P}\right)\left[1-(1-r)^{k}\right].\label{eq:39 ancilla heat}
\end{align}       

These equations show consistently that whenever the sample is heated up the ancilla is cooled down and vice versa. The plots in Fig. S2 also corroborate this behavior. To obtain these graphs, we consider fixed physical parameters identical to those of Fig. 2
in the main text. Specifically, for Fig. 4(a) the prior temperature
is $\widetilde{T}=\varepsilon_{s}/4$ and we consider two sample temperatures
$T_{s}=\varepsilon_{s}/4.5$ and $T_{s}=\varepsilon_{s}/3.5$. Figure
4(b) shows plots for the prior temperature is $\widetilde{T}=\varepsilon_{s}/10$
and two sample temperatures $T_{s}=\varepsilon_{s}/10.5$ and $T_{s}=\varepsilon_{s}/9.5$.  

\begin{figure}
\begin{centering}
\includegraphics[scale=0.34]{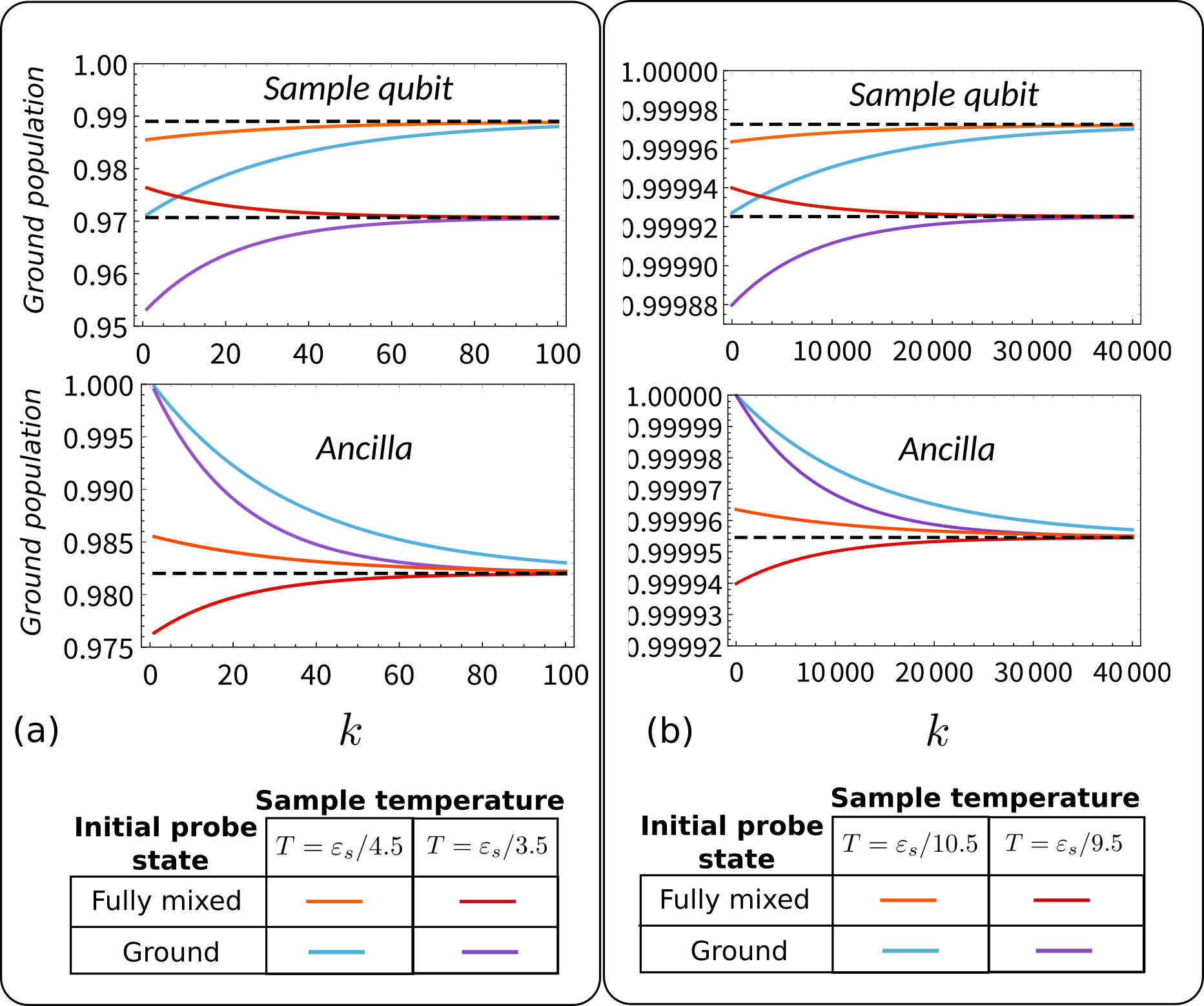}
\caption{Perturbation of the sample and the ancilla, generated by the joint interaction with the probe (color conding in the bottom charts). The two upper plots show the ground population of the $(k+1)$th sample qubit and the two lower plots show the ground population of the ancilla, after the $(k+1)$th interaction. The black dashed lines correspond to the steady-state values. Fixed parameters in (a) and (b) are respectively the same in Figs. 2(a) and 2(b) of the main text. Thus, the prior temperatures are $\widetilde{T}=\varepsilon_{s}/4$ for (a) and $\widetilde{T}=\varepsilon_{s}/10$ for (b).}
\end{centering}
\end{figure}

The plots in Fig. S1 depict the ground population of the sample qubits
and the ancilla as a function of the number of interactions with the
probe. For $k$ sufficiently large, the probe approaches its steady
state and hence the perturbations that it causes become negligible.
As a consequence, the populations approach their unperturbed values
given by the black dashed lines. 

From these plots we can draw the following insights: 
\begin{enumerate}
\item Preparing the probe in the fully mixed state generally causes less
perturbation, as indicated by the red and orange curves. 
\item If the probe starts in its ground state (blue and purple curves),
the ancilla is always cooled down and the sample is always \textit{heated
up}. On the other hand, the red curves show that the opposite effect
can occur if the probe starts in the fully mixed state. Since cooling
back the sample is naturally more difficult than reheating it, this
shows that to minimize the disturbance of the sample it is preferable
to initialize the probe in the fully mixed state. 
\item For larger temperatures (Fig. S1(a)) the perturbation of both the
sample and the ancilla is larger. While the corresponding population
variations are between orders of $10^{-3}$ and $10^{-2}$, the order
of the variations in Fig. S1(b) is $10^{-5}$. This is noteworthy
because it suggests that colder samples are subject to less (absolute)
perturbation. 
\end{enumerate}

\section{Comparison between the transient SNR (Eq. (\ref{eq:11 transient SNR})
in the main text) and the SNR for an optimal probe-sample interaction }

Reference \cite{hovhannisyan2021optimal} studies optimal POVMs for coarse-grained thermometry, where the number of outcomes of the POVM is smaller than the dimension of the measured sample. In particular, a generic two-outcome POVM can be implemented by letting the sample to undergo an arbitrary interaction (possibly involving ancillas) with a two-level probe. Given this equivalence between coarse-grained and probe-based thermometry, we can apply the results of \cite{hovhannisyan2021optimal} to compare the performance of the thermometric machine with the maximum SNR that results from such an optimal interaction. This comparison is performed in Fig. S2, for fixed parameters $\varepsilon_{s}/T=8$ and $\varepsilon_{s}/\widetilde{T}=7$. 

The red solid curve in Fig. S2 shows the ratio between the optimal probe SNR and the sample SNR (obtained via an energy measurement of $k$ qubits). This ratio tends asymptotically towards the value $\sqrt{2/\pi}$, as indicated by the black dashed line. On the other hand, the blue solid curve depicts the ratio between the transient SNR of our machine's probe and the sample SNR. For the considered parameters $k\sim6000$ interactions allow to approximate very well the steady-state SNR $\sim3.54$. Clearly, more interactions cannot increase this value and therefore the blue curve is monotonically decreasing. However, it is remarkable that in the interval $k\in[1,6000]$ the machine produces an SNR that is at least $\sim38\%$ of the optimal one, even though in the worst case scenario this would require a $(k+1)$-body interaction. 

\begin{figure}
\includegraphics[scale=0.6]{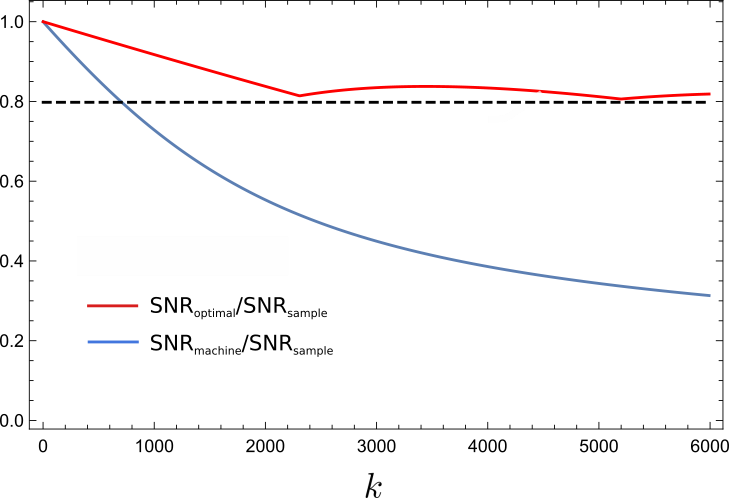}
\caption{Comparison between the transient SNR achieved with the machine and the maximum probe SNR corresponding to an optimal probe-sample interaction. Fixed parameters are $\varepsilon_{s}/T=8$ and $\varepsilon_{s}/\widetilde{T}=7$. In all the cases the SNRs are ``normalized'' with respect to the sample SNR and computed for a single measurement $M=1$.}
\end{figure}

\end{document}